\documentclass[
onecolumn,11pt,
nofootinbib]{revtex4}

\usepackage[utf8]{inputenc}

\usepackage{amsmath,amssymb,mathtools}
\usepackage{graphicx}

\newcommand{\C}{\mathbb{C}} %

\newcommand{\one}{\mathbb{I}} %

\renewcommand{\Re}{\operatorname{Re}} 
\renewcommand{\Im}{\operatorname{Im}} 

\newcommand{\p}{\partial} 

\DeclareMathOperator{\tr}{Tr} 
\DeclareMathOperator{\diag}{diag} 
\DeclareMathOperator{\str}{Str} 
\DeclareMathOperator{\sign}{sign} 

\newcommand{\cL}{\mathcal L}
\newcommand{\cN}{\mathcal N}

\newcommand{\nn}{\nonumber}

\let\olddet\det
\renewcommand{\det}{\olddet\nolimits} 

\renewcommand{\phi}{\varphi}
\renewcommand{\epsilon}{\varepsilon}

\DeclarePairedDelimiter{\abs}{\lvert}{\rvert} 
\DeclarePairedDelimiter{\inner}{\langle}{\rangle} 

\begin{document}

\title{Baryon Number Dirac Spectrum in QCD}

\author{J.~R.~Ipsen~and K.~Splittorff}
\affiliation{
Discovery Center, Niels Bohr Institute, University of Copenhagen, Blegdamsvej 17, DK-2100 Copenhagen \O, Denmark.}

\begin{abstract}
The relation between the baryon number in QCD at nonzero chemical potential and the spectral density of the baryon number Dirac operator, $\gamma_0(D+m)$, is examined. We show that extreme oscillations of the spectral density, caused by the QCD sign problem, are essential for the formation of the average baryon number when $\mu>m_\pi/2$. We compute the oscillating region of the spectral density using chiral perturbation theory. The extreme oscillations have a microscopic period and are resolved using random matrix theory.
\end{abstract}

\maketitle

\newpage

\section{Introduction}
\label{sec:intro}

\noindent
Lattice QCD \cite{LatticeQCD} is our best nonperturbative tool to study the phase structure of strongly interacting matter as a function of temperature. At low temperature the formation of the chiral condensate, which signals the spontaneous breaking of chiral symmetry, can be read of from the density of eigenvalues of the lattice Dirac operator at the origin. This useful link between the chiral condensate and the eigenvalues of the Dirac operator is known as the Banks-Casher relation \cite{Banks:1980}.

At nonzero chemical potential, $\mu$, the Monte Carlo method, which forms the basis of Lattice QCD simulations, is hampered by {\sl the sign problem}: The fermion determinant of the Dirac operator, and consequently the measure on which we would like to perform Monte Carlo sampling, takes complex values, see for example the reviews \cite{sign-prob-rev}. 
Also the Banks-Casher relation \cite{Banks:1980} breaks down at nonzero chemical potential. The new relation \cite{Osborn:2005,Osborn:2008b}, which replaces the Banks-Casher relation, explicitly shows the intimate connection between the sign problem and the spontaneous breaking of chiral symmetry: it is a region with extreme oscillations of the eigenvalue density of the Dirac operator \cite{Osborn:2004,Akemann:2004}, caused by the sign problem, which leads to the discontinuity of the chiral condensate in the chiral limit at nonzero chemical potential. The oscillations have a period of order the inverse volume and an amplitude which is exponentially large in the volume. The region within which the oscillations take part can be computed using the mean field approximation in chiral perturbation theory \cite{Osborn:2008} but as the period of the oscillations are on the microscopic scale, set by the inverse volume, one needs the exact microscopic eigenvalue density \cite{Osborn:2004,Akemann:2004} in order to resolve them. For a review see e.g.~\cite{XQCD}.

The new relation \cite{Osborn:2005,Osborn:2008b}, which replaces the Banks-Casher relation, not only give a direct insight in the way the chiral condensate is formed in unquenched QCD at nonzero chemical potential. It also solves the long standing problem of how the eigenvalues of the Dirac operator, $D$, can have a strong dependence on the chemical potential while at low temperature the partition function is independent of the chemical potential when $\mu$ is less than a third of the nuclon mass (in \cite{Cohen:2003} this was coined the Silver Blaze problem). The relation between a region of the eigenvalue density with extreme oscillations and the spontaneous breaking of chiral symmetry has also been established in 1dQCD \cite{Ravagli:2007} as well as in two color QCD with non degenerate quark masses \cite{Akemann:2011}.

Here we show that the mechanism behind the new relation \cite{Osborn:2005,Osborn:2008b} is not restricted to the chiral condensate and the spectral density of $D$: It is the exact same mechanism that links the baryon number density to the spectral density of baryon number Dirac operator, $\gamma_0(D+m)$.

The average baryon number is given by
\begin{equation}
\label{nBdef}
Vn_B(\mu)=\frac{d}{d\mu}\log Z(\mu)=\inner[\Big]{\tr\frac{1}{D_0(m)+\mu}},
\end{equation}
where the baryon number Dirac operator is, $D_0(m)\equiv \gamma_0(D+m)$. The average trace in (\ref{nBdef}) can be expressed as the integral
\begin{equation}
Vn_B(\mu)=\int_\C \frac{d^2\lambda}{\lambda+\mu}\rho(\lambda,\lambda^\ast),
\label{intro:nB}
\end{equation}
where $\lambda$ are the eigenvalues of the baryon number Dirac operator and $\rho$ is the spectral density of the baryon number Dirac operator. 

We will compute the eigenvalue density of $D_0(m)$ and demonstrate that it is again a strongly oscillating region of the unquenched eigenvalue density which ensures the correct physical behavior of the average baryon density. In order to establish this result we will investigate the structure of the spectral density of the baryon number Dirac spectrum using chiral perturbation theory ($\chi$PT) and random matrix theory (RMT).

In quenched simulations a strongly oscillating region of the eigenvalue density is not possible, since the quenched theory is free from the sign problem. In such simulations the baryon number density therefore becomes nonzero when the chemical potential reaches half the pion mass rather than at a third of the baryon mass as expected for unquenched QCD. This early onset of the baryon density in the quenched theory has been observed on the lattice \cite{Barbour:1986,Gibbs:1986} and in a $U(1)$ model \cite{Gocksch:1988}. It was also understood analytically \cite{Stephanov:1996} within a random matrix model (RMM).

Before we turn to the actual computation of the eigenvalue density of the baryon number Dirac operator we give, in section~\ref{sec:example}, a simple example which illustrates the mechanism involving a strongly oscillating density. Then in section~\ref{sec:mf} we will use a mean field approach to describe the phase diagram of the spectrum. Finally, in section~\ref{sec:rmt}, we will use a RMM to show that the extreme oscillations of the microscopic eigenvalue density lead to the expected behavior of the average baryon number density.

\section{An Example with Oscillating Regions}
\label{sec:example}

In this section we will give a simple example to illustrate the point that oscillations with a microscopic wavelength can remove the unphysical early onset of the average baryon number at $\mu=m_\pi/2$ observed in the quenched theory. 

First, let us consider the nonoscillating eigenvalue density corresponding to the quenched case  
\begin{equation}
\rho^{\rm Q}_\text{Ex}(x,y,m_\pi)=\frac{1}{\pi}F_\pi^2V\Big(1+\frac{3m_\pi^4}{16x^4}\Big)\theta(\abs x-m_\pi/2),
\label{example:rho}
\end{equation}
where $x$ and $y$ are the real and imaginary part of the eigenvalue $\lambda$, see figure \ref{fig:phases_q}. This spectral density results, through \eqref{intro:nB},  in a baryon density, which is zero for $\mu<m_\pi/2$ and nonzero for $\mu>m_\pi/2$,
\begin{equation}
\begin{split}
Vn^{\text{Ex,Q}}_B(\mu)
&=\int dx\int dy \frac{1}{x+iy+\mu}\frac{1}{\pi}F_\pi^2V\Big(1+\frac{3m_\pi^4}{16x^4}\Big)\theta(\abs{x}-m_\pi/2) \\
&=F_\pi^2V\int dx \sign(x+\mu) \Big(1+\frac{3m_\pi^4}{16x^4}\Big) \theta(\abs{x}-m_\pi/2) \\
&=2F_\pi^2V\Big(\mu-\frac{m_\pi^4}{16\mu^3}\Big) \theta(\abs\mu-m_\pi/2).
\end{split}
\end{equation}
This is like the observed quenched baryon density, see figure~\ref{fig:densities_rho_n}. Let us now illustrate the main point: The unquenched average baryon number is zero also when $\mu$ is in the range between half the pion mass and a third of the nucleon mass because of two strongly oscillating regions. In order to see how this works we extend the example as
\begin{equation}
\rho^{\rm UnQ}_\text{Ex}(x,y,\mu,m_\pi)=\rho^{\rm Q}_\text{Ex}(x,y,m_\pi)(1-e^{F_\pi^2V\,U(x,y,\mu,m_\pi)}), 
\end{equation}
with
\begin{equation}
U(x,y,\mu,m)=-(y+i\sign(x)(\abs{x}+\abs{\mu}))^2-m_\pi^2-(4x^2-m_\pi^2)\theta(\abs x-m_\pi/2),
\end{equation}
and where $V$ is the four volume. Note that the amplitude of the oscillations grows exponentially with $V$, and that the period of the oscillations (determined by $e^{-2iy\sign x(\abs x+\abs\mu)F_\pi^2V}$) is of order $1/V$.

The contribution to the average baryon density from the oscillating part of the eigenvalue density can be evaluated by a saddle point integration, in the large $V$ limit. The contour along the real $y$ axis is deformed to go through the saddle point $y=-i\sign(x)(\abs{x}+\abs{\mu})$. If the deformation passes the pole at $y=i(x+\mu)$, then we add a contribution for integration around the pole. We obtain
\begin{equation}
\int_{-\infty}^\infty dy\frac{e^{F_\pi^2 V\, U(x,y,\mu,m)}}{x+iy+\mu}
=2\pi[\theta(\mu)\theta(-x)\theta(x+\mu)-\theta(-\mu)\theta(x)\theta(-x-\mu)].
\end{equation}
It follows that the average baryon number is zero also for $\mu>m_\pi/2$
\begin{equation}
Vn^{\text{Ex,UnQ}}_B(\mu)=\int \frac{dxdy}{x+iy+\mu}\rho^{\rm UnQ}_\text{Ex}(x,y,\mu,m_\pi)=0.
\end{equation}
What we have learned from this example is that bounded oscillating regions can remove the unphysical early onset of the average baryon number at $\mu=m_\pi/2$. Below we show, using $\chi$PT and RMT, that this mechanism is realized in unquenched QCD at nonzero $\mu$.

\section{The boundaries of the oscillating regions}
\label{sec:mf}

In this section we will investigate the phase diagram of the baryon number Dirac spectrum by means of $\chi$PT at mean field level. Our aim is to calculate the boundaries of the oscillating regions. To do this we employ the replica trick (see e.g.~\cite{Damgaard:2000}), where $n$ replica pairs consisting of quarks and conjugate quarks has been introduced in the partition function. The replica partition function is given by the ensemble average of $N_f$ quarks and $n$ replica pairs,
\begin{equation}
Z^{N_f,n}(\lambda,\lambda^\ast,\mu,m)=
\inner[\Big]{\det^n(D_0(m)+\lambda)\det^n(D_0(m)-\lambda^\ast){\det}^{N_f}(D_0(m)+\mu)}.
\label{mf:Z}
\end{equation}
The spectral density of the baryon number Dirac operator is obtained from the replica partition function by (see \cite{Feinberg:1997} for a detailed explanation of how the density results from the derivatives)
\begin{equation}
\rho^{N_f}(\lambda,\lambda^\ast,\mu,m)=\lim_{n\to 0}\frac{1}{n}\p_\lambda\p_{\lambda^\ast}
\log Z^{N_f,n}(\lambda,\lambda^\ast,\mu,m).
\end{equation}
Note that the chemical potential of the replica quarks, $\lambda$, corresponds to the eigenvalues of $D_0(m)$.

We will examine the partition function $Z^{N_f,n}(\lambda,\lambda^\ast,\mu,m)$ to leading order in chiral perturbation theory. The chiral Lagrangian to leading order is given by~\cite{Gasser:1984,Gasser:1985}
\begin{equation}
\cL_\text{eff}=-\frac{F^2_\pi}{4}\tr D_\nu UD_\nu U^\dagger-\frac{\Sigma}{2}\tr M(U+U^\dagger),
\label{mf:rho}
\end{equation}
where $\Sigma$ is the magnitude of the chiral condensate and $F_\pi$ is the pion decay constant. The chemical potential enters the QCD partition function as en external vector field, $B_\nu=\delta_{\nu0}B$. To respect local invariance, $B_\nu$ can only appear in the covariant derivative, see e.g.~\cite{Pich:1998,Son:2000}
\begin{equation}
D_\nu U=\p_\nu U+i[U,B_\nu].
\label{mf:DU}
\end{equation}
In the case where we have $N_f$ quarks and $n$ replica pairs all with mass $m$ and with chemical potential $\mu$ and $\lambda$, respectively, then we have that
\begin{equation}
U\in SU(2n+N_f),\quad M=m\one_{2n+N_f}\quad \text{and}\quad B=\diag(\lambda\one_n,-\lambda^\ast\one_n,\mu\one_{N_f}).
\label{mf:UMB}
\end{equation}

\subsection{The Quenched Theory}
\label{sec:q_mf}

\begin{figure}[htbp]
\begin{minipage}[c]{.99\textwidth}
\includegraphics{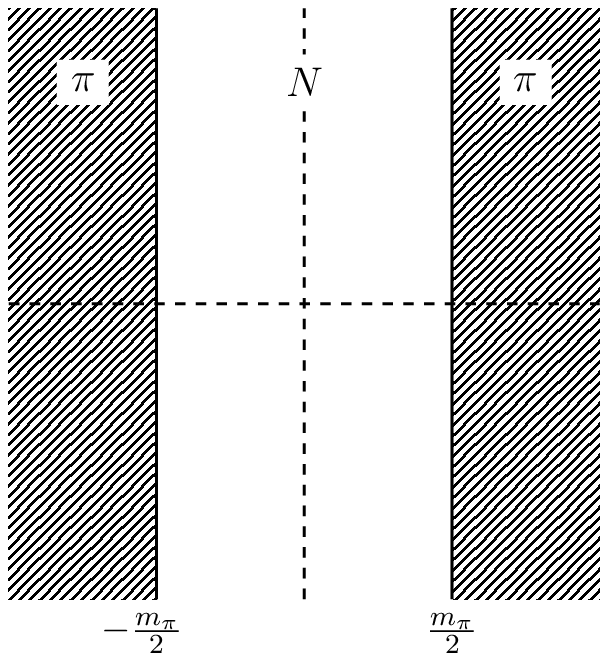}
\hspace{.1\textwidth}
\includegraphics{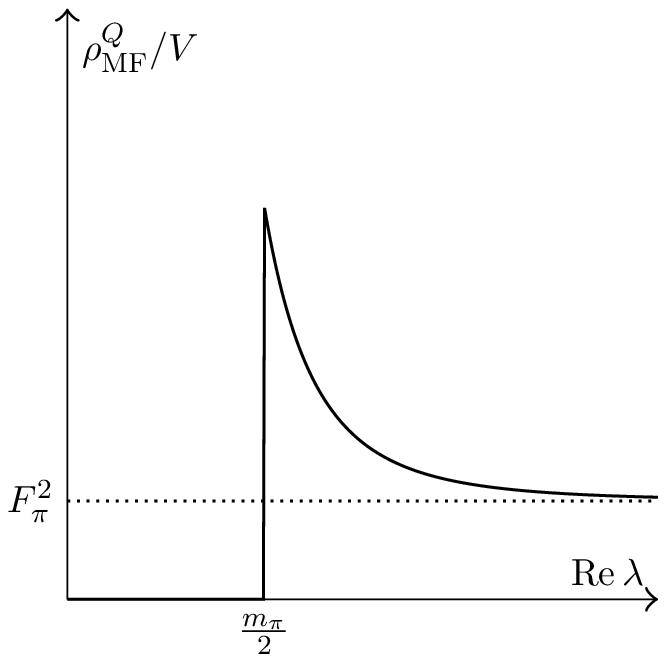}
\end{minipage}
\caption{\textbf{Left panel:} The phase diagram of the quenched baryon number Dirac spectrum in the complex eigenvalue $\lambda$-plane. The diagram has a strip of zero spectral density and width $m_\pi$ centered around the imaginary $\lambda$ axis. The normal phase is labeled by $N$ and $\pi$ refers to pion condensed phase. The eigenvalue density is nonzero in the pion condensed phase.  \textbf{Right panel:} Plot of the quenched spectral density of the baryon number Dirac operator as a function of $\Re\lambda$ (the density is independent of $\Im\lambda$). If $m=0$, the pion is massless and the spectral density is represented by the dotted line.}
\label{fig:phases_q}
\end{figure}

Here we look at the quenched theory, where $N_f=0$, and the partition function~\eqref{mf:Z} is given by the quenched ensemble average of $n$ replica pairs,
\begin{equation}
Z^n(\lambda,\lambda^\ast,m)=\inner[\big]{\det^n(D_0(m)+\lambda)\det^n(D_0(m)-\lambda^\ast)},
\label{q_mf:Z}
\end{equation}
where $m$ is the mass of the replica quarks and $\lambda$ is the complex chemical potential. The spectral density of the baryon number Dirac operator is obtained as
\begin{equation}
\rho^{N_f=0}(\lambda,\lambda^\ast,m)=\lim_{n\to 0}\frac{1}{n}\p_\lambda\p_{\lambda^\ast}
\log Z^n(\lambda,\lambda^\ast,m).
\label{q_mf:rho}
\end{equation}
In the mean field limit the free energy has a trivial linear dependence on $n$. For this reason we can restrict ourselves to $n=1$ in the following calculation. From~\eqref{q_mf:Z} with $n=1$ one sees that the chemical potential of replica quarks corresponds to isospin chemical potential for real $\lambda$. At low temperature and $\mu>m_\pi/2$ there will be a condensate of quarks and conjugate quarks. One can interpret the quarks and conjugate quarks as ``up'' and a ``down'' quarks, and for this reason we will refer to the condensate of replica quarks as the pion condensate.

\begin{figure}[htbp]
\includegraphics{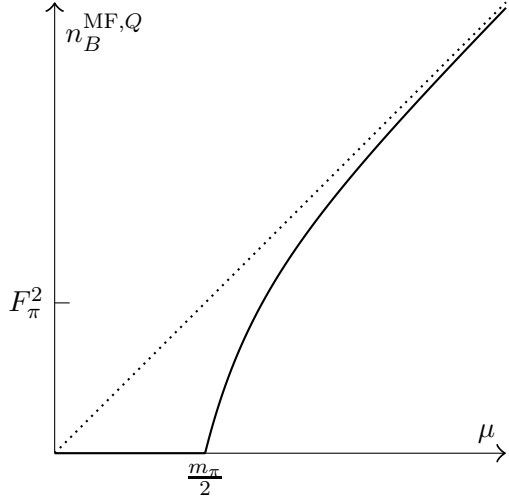}
\caption{A plot of the quenched mean field baryon number density (solid line). If $m=0$, then the pions are massless and baryon density is represented by the dotted line.}
\label{fig:densities_rho_n}
\end{figure}

The partition function $Z^{n=1}(\lambda,\lambda^\ast,m)$ depends on the chiral Lagrangian given in~\eqref{mf:rho} and~\eqref{mf:DU} with $U\in SU(2)$, $M=\diag(m,m)$ and $B=\diag(\lambda,-\lambda^\ast)$. The mean field phase diagram, however, only depends on the static part of the Lagrangian
\begin{equation}
\cL_\text{stat}=\frac{F^2_\pi}{4}\tr [U,B][U^\dagger,B]-\frac{\Sigma}{2}\tr M(U+U^\dagger).
\label{q_mf:L}
\end{equation}
To find the mean field structure we use the ansatz, 
\begin{equation}
U=\begin{pmatrix}\cos\alpha & \sin\alpha \\ -\sin\alpha &\cos\alpha\end{pmatrix},
\end{equation}
which results in an effective Lagrangian
\begin{equation}
\cL_\text{stat}=-\frac{F^2_\pi}{2}(\lambda+\lambda^\ast)^2\sin^2\alpha-2m\Sigma\cos\alpha.
\end{equation}
The minima of this Lagrangian are easily found to be at $\sin\alpha=0$ and $\cos\alpha=2m\Sigma/F^2_\pi(\lambda+\lambda^\ast)^2$. This yields the free energies
\begin{equation}
\label{LstatQ}
\begin{split}
V\cL^N_\text{stat}&=-2m\Sigma V, \\
V\cL^\pi_\text{stat}&=-\frac{F^2_\pi V}{2}(\lambda+\lambda^\ast)^2-\frac{2m^2\Sigma^2V^2}{F^2_\pi V(\lambda+\lambda^\ast)^2},
\end{split}
\end{equation}
for the normal and pion phase, respectively. The boundary between the two phases is given by
\begin{equation}
(\Re\lambda)^2=\frac{2m\Sigma}{4F^2_\pi}=\Big(\frac{m_\pi}{2}\Big)^2,
\end{equation}
where we used the Gell-Mann-Oakes-Renner relation for the Goldstone mass~\cite{GellMann:1968}, $m^2_\pi=2m\Sigma/F^2_\pi$. 
Using~\eqref{q_mf:rho} we find that the spectral density is zero in the normal phase and given by
\begin{equation}
\rho_\pi(\lambda,\lambda^\ast,m)=F^2_\pi V+\frac{12m^2\Sigma^2V}{F^2_\pi(\lambda+\lambda^\ast)^4}
=F^2_\pi V\Big(1+\frac{3m^4_\pi}{(\lambda+\lambda^\ast)^4}\Big),
\end{equation}
in the pion phase.  This structure is consistent with that obtained from a RMM in \cite{Halasz:1999,Halasz:1997}.  Note that the phases of the replicated partition function are directly linked to the behavior of the eigenvalue density, hence we speak of the phase diagram of the baryon number Dirac spectrum. 
In terms of the microscopic parameters $\hat m=m\Sigma V$ and $\hat \lambda=\lambda F_\pi\sqrt V$ we have
\begin{equation}
\rho_\pi(\hat x,\hat m)=F^2_\pi V\Big(1+\frac{3}{4}\frac{\hat m^2}{\hat x^4}\Big),
\end{equation}
where $x\equiv \Re\lambda$. A plot of the quenched phase diagram and the spectral density is given in figure~\ref{fig:phases_q}.
%

We will now look at the quenched average baryon number,
\begin{equation}
Vn^{\text{MF},Q}_B(\mu,m)=\int_\C d^2\lambda \frac{\rho^Q_\text{MF}(\lambda,\lambda^\ast,m)}{\lambda+\mu},
\end{equation}
where $\lambda$ is the eigenvalues of $D_0(m)$, and the quenched mean field spectral density is given by
\begin{equation}
\rho^Q_\text{MF}(\lambda,\lambda^\ast,m)=
F^2_\pi V\Big(1+\frac{3}{4}\frac{\hat m^2}{\hat x^4}\Big)
\theta(\abs{x}-m_\pi/2).
\label{rho_q_mf}
\end{equation}
This is precisely the quenched density we used in the example and again integrating over the real and imaginary part of $\lambda$ leads to
\begin{equation}
\begin{split}
Vn^{\text{MF},Q}_B(\hat\mu)
&=F_\pi^2V\int d\hat x\int d\hat y \frac{1}{\hat x+i\hat y+\hat\mu}
\Big(1+\frac{3\hat m^2}{4\hat x^4}\Big)\theta(\abs{x}-m_\pi/2) \\
&=F_\pi^2V\Big(\hat\mu-\frac{\hat m^2}{4\hat\mu^3}\Big) \theta(\abs{\mu}-m_\pi/2).
\end{split}
\end{equation}
On figure~\ref{fig:densities_rho_n} one sees that the quenched theory predicts a nonzero baryon density for $\mu>m_\pi/2$ as mentioned in the introduction. For $m=0$ the mass of the Goldstone bosons are zero. In this case the spectral density is a constant and the average baryon density is linear (see figure~\ref{fig:densities_rho_n}).

\subsection{The Unquenched Theory}
\label{sec:unq_mf}

\begin{figure}[htbp]
\includegraphics{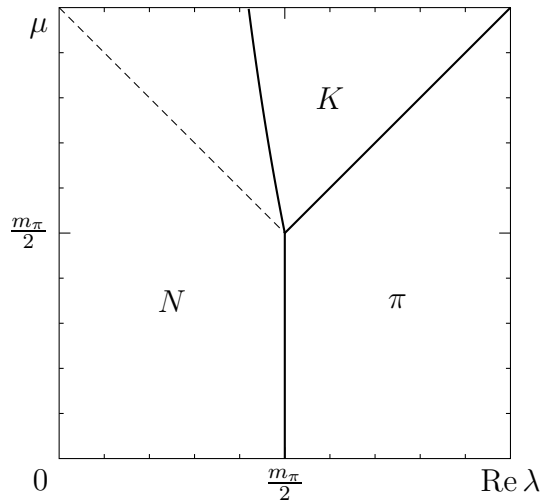}
\caption{Unquenched phase diagram in the $\Re\lambda$---$\mu$ plane. The labels $N$, $\pi$ and $K$ refer to the normal, pion and kaon condensed phases. The dashed line markes the boundary between the normal and kaon phase. To the right of the full line in the kaon phase the eigenvalue density is expected to be strongly oscillating.
Note that the kaon condensed phase is introduced in the spectrum at $\mu=m_\pi/2$, which is exactly where the unphysical behavior of the quenched theory occur.}
\label{fig:phases}
\end{figure}

\begin{figure}[htbp]
\begin{minipage}[t]{.99\textwidth}
\includegraphics{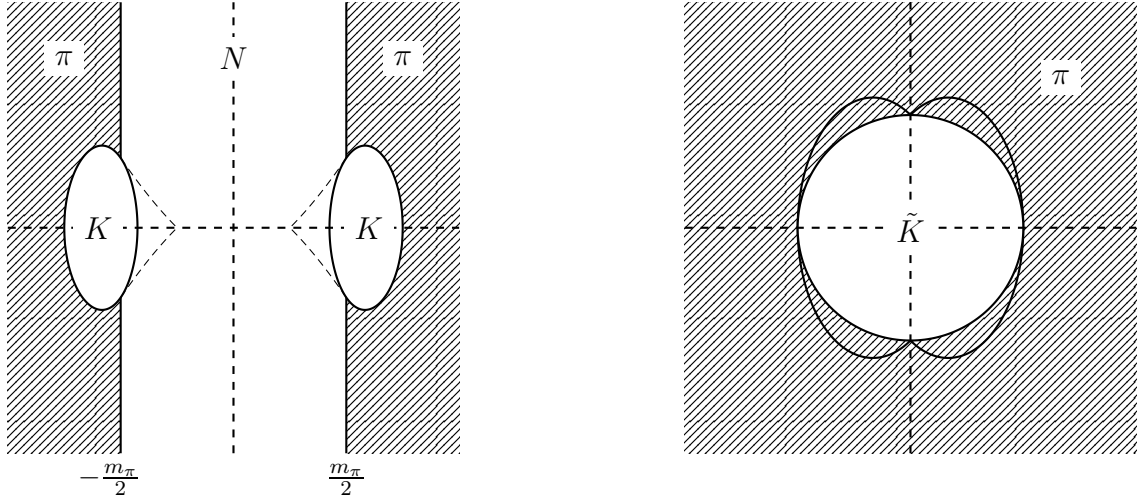}
\end{minipage}
\caption{\textbf{Left panel:} Phase diagram for the spectrum of the baryon number Dirac operator with $\mu>m_\pi/2$. The labels $N$, $\pi$ and $K$ refer to the normal, pion and kaon condensed phases. The density is expected to be strongly oscillating within the regions marked by K, which crosses the real axis at $x=\pm\tfrac{1}{2}(-\mu+\sqrt{\mu^2+2m_\pi^2})$ and $x=\pm\mu$. \textbf{Right panel:} Phase diagram for the spectrum of the baryon number Dirac operator with massless quarks ($m=0$). The circular region contain strong oscillations (see section~\ref{sec:rmt_m=0}) and has radius $\mu$.}
\label{fig:phases_unq}
\end{figure}

We now consider the unquenched case. Just as in the queched theory the phases of the replicated partition function are directly linked to the behavior of the eigenvalue density of the baryon number Dirac opartor.  
In mean field $\chi$PT the boundaries between the phases in the unquenched phase diagram of the baryon number Dirac spectrum are independent of the number of quark flavors and replicas. For this reason we will concentrate on the case $n=1$ and $N_f=1$. The replica partition function is given by
\begin{gather}
Z^{N_f=1,n=1}(\lambda,\lambda^\ast,\mu,m)=
\inner[\Big]{\det(D_0(m)+\lambda)\det(D_0(m)-\lambda^\ast)\det(D_0(m)+\mu)},
\end{gather}
where $m$ is the mass of both ordinary quarks and replicas, $\mu$ is the ordinary chemical potential and $\lambda$ is the replica chemical potential. Because of the similarity with the phase diagram of the $(u,d,s)$ quark triplet~\cite{Kogut:2001}, we will refer to the condensate of ordinary quarks with replicas as the kaon condensate. The masses of the Goldstone bosons (pion and kaon) are in this case equal
\begin{equation}
m^2_\pi=m^2_K=2m\Sigma/F^2_\pi.
\end{equation}
The static chiral Lagrangian to lowest order is the same as in equation~\eqref{q_mf:L} but with $M=m\one_3$, $B=\diag(\lambda,-\lambda^\ast,\mu)$ and of $U\in SU(3)$. In this case we make an ansatz for the Goldstone fields given by
\begin{equation}
U=R_1(\alpha)R_3(\beta),
\end{equation}
where $R_j(\alpha)$ is a rotation by $\alpha$ about the $j$ axis. This implies that the static Lagrangian becomes
\begin{multline}
\cL_\text{stat}=-\frac{F^2_\pi}{4}
\Big[\Big(2(\lambda^\ast+\mu)^2+(\lambda+\lambda^\ast)(\lambda^\ast-\lambda+2\mu)\sin^2\beta\Big)\sin^2\alpha
+(\lambda+\lambda^\ast)^2\sin^2\beta(1+\cos^2\alpha)\Big] \\
-m\Sigma(\cos\beta+\cos\alpha(1+\cos\beta)).
\end{multline}
We will focus on the minima with $\mu>0$ and $x=\Re\lambda>0$, since the rest of the phase diagram is given by symmetry. For $\alpha=0$ there are two minima:
\begin{equation}
\sin\beta=0\qquad {\rm and} \qquad \cos\beta=2m\Sigma/F_\pi^2(\lambda+\lambda^\ast)^2.
\end{equation}
These minima give the free energy of the normal and pion condensed phases
\begin{subequations}
\label{unq_mf:F}
\begin{equation}
\begin{split}
V\cL_\text{stat}^N&=-3m\Sigma V, \\
V\cL_\text{stat}^\pi&=-m\Sigma V-\frac{F^2_\pi V(\lambda+\lambda^\ast)^2}{2}-\frac{2m^2\Sigma^2 V^2}{F^2_\pi V(\lambda+\lambda^\ast)^2},
\quad(x^2\geq m\Sigma/2F^2_\pi).
\end{split}
\end{equation}
Note that the $\lambda$ and $\lambda^*$ dependence at these minima are exactly as in the quenched case, cf.~(\ref{LstatQ}). When these minima are dominant the density is therefore identical to that found in the quenched case. 

The unquenched theory has a third minimum given by $\alpha=\pi/2$ and $\cos\beta=2m\Sigma/F^2_\pi(\lambda^\ast+\mu)^2$, with implies the free energy
\begin{align}
V\cL_\text{stat}^K=-m\Sigma V-\frac{F^2_\pi V(\mu+\lambda^\ast)^2}{2}-\frac{2m^2\Sigma^2V^2}{F^2_\pi V(\mu+\lambda^\ast)^2}.\quad
\end{align}
\end{subequations}
The phase boundaries occur where the real part of the free energy of two different phases are equal. The boundary between the normal and pion condensed phase is given by half the pion mass, just as in the quenched case. The formal expression of the normal-kaon and the pion-kaon boundaries are both given by a cubic equation in $\hat y^2\equiv\Re\hat\lambda^2$ with one real root and two complex conjugate roots. The structure of the kaon boundary can be seen on figure~\ref{fig:phases} and~\ref{fig:phases_unq}. For $\mu<m_\pi/2$ the unquenched phase diagram is identical to the quenched case (see~figure~\ref{fig:phases_q}). This is exactly what is expected, since the quenched theory trivially gives correct predictions for the average baryon density as long as $\mu<m_\pi/2$.  For $\mu>m_\pi/2$ the kaon phase appears in the unquenched spectral phase diagram, see Figure~\ref{fig:phases}. We expect that the eigenvalue density is strongly oscillating within the kaon region and that these oscillations cure the unphysical behavior of the quenched baryon density, just as in the example in section \ref{sec:example}. 
A diagram which shows the phase boundaries for $\mu>m_\pi/2$ of the kaon condensed phase are given in figure~\ref{fig:phases_unq}.
The details of the oscillations of the eigenvalue density are not resolved at the mean field level because the period of the oscillations is of order $1/V$. Rather at mean field level one finds simply $\rho=0$ in the kaon phase. In section~\ref{sec:rmt_m=0} we will compute the microscopic eigenvalue density using a RMM and show that the kaon condensed phase is indeed a strongly oscillating region, which exactly cancels the unphysical early onset of the average baryon density observed in the quenched case.
\subsection{The Massless Case}

In the limit of zero quark mass an additional saddelpoint is present. Here we analyze this case. In section~\ref{sec:rmt_m=0} will compute the microscopic spectral density using a RMM of massless quarks, and the results of the present subsection will allow us to compare directly between $\chi$PT and the RMM.

At $m=0$ the static part of the chiral Lagrangian~\eqref{q_mf:L} is given by
\begin{multline}
\cL_\text{stat}=-\frac{F^2_\pi}{4}
\Big[\Big(2(\lambda^\ast+\mu)^2+(\lambda+\lambda^\ast)(\lambda^\ast-\lambda+2\mu)\sin^2\beta\Big)\sin^2\alpha
+(\lambda+\lambda^\ast)^2\sin^2\beta(1+\cos^2\alpha)\Big],
\end{multline}
which has four minima. The minima and corresponding free energies are
\begin{center}
\begin{tabular}{c @{\qquad\qquad} c @{\qquad} c @{\qquad\qquad} c}
Phase & $\alpha$         &$\beta$             &Free energy, $V\cL(\alpha,\beta)$                 \\
\hline
$N$   & $0$             &$0$                  & $0$                                              \\
$\pi$ & $0$             &$\frac{\pi}{2}$      & $-2\hat x^2$                                     \\
$K$   & $\frac{\pi}{2}$ &$0$                  & $-\tfrac{1}{2}(\hat x-i\hat y+\hat \mu)^2$       \\
$\tilde{K}$ & $\frac{\pi}{2}$ &$\frac{\pi}{2}$      & $\tfrac{1}{2}(-3\hat x^2+(\hat y+i\hat \mu)^2)$  
\end{tabular}
\end{center}
The first three free energies can be found from the massive mean field results~\eqref{unq_mf:F} by setting $m=0$; the additional minimum of the massless theory exist since all minima trivially fulfill the constraints $\abs{\cos\alpha}\leq1$ and $\abs{\cos\beta}\leq1$.

The free energy of the normal phase is never a global minimum in the theory of massless quarks. The phase diagram of the quenched theory is therefore everywhere dominated by the condensate of the (massless) pions. Unquenching introduces condensation between replica and physical flavors. The phase boundary between these phases and the pion condensed phase can be seen in the right hand panel of figure~\ref{fig:phases_unq} as the solid lines. The circular region is given by the boundary between pion and the $\tilde K$ phase, and in section~\ref{sec:rmt_m=0} we will show that this phase is dominated by strong oscillations, which exactly cancels the unphysical behavior of the quenched theory. In case of massless quarks the kaon condensed phase is a false minimum coming from the incomplete description of the fermionic replica trick. For a critical discussion of the replica method see \cite{Verbaarschot:1985,Zirnbauer:2003}. 
The problem becomes explicit within the supersymmetric technique (see \cite{Efetov,Toublan:1999} for an introduction) where one obtains the density from
\begin{equation}
\rho^{N_f}(\lambda,\lambda^\ast,\mu,m)=\lim_{\lambda'\to\lambda}\p_\lambda\p_{\lambda^\ast}
\log Z^{N_f+2|2}(\lambda,\lambda^\ast,\lambda',\lambda'^\ast,\mu,m),
\end{equation}
where 
\begin{equation}
Z^{N_f+2|2}(\lambda,\lambda^\ast,\lambda',\lambda'^\ast,\mu,m)=
\inner*{\frac{\det(D_0(m)+\lambda)\det(D_0(m)-\lambda^\ast)}{\det(D_0(m)+\lambda')\det(D_0(m)-\lambda'^\ast)} \ {\det}^{N_f}(D_0(m)+\mu)}.
\end{equation}
The presence of both fermionic and bosonic determinants leads to both ordinary and fermionic Goldstone modes in the effective $\chi$PT formulation. Integrating out the noncommuting variables gives a partially quenched partition function of the form
\begin{equation}
Z^{N_f+2|2}=\int d\mu_F\int d\mu_B\, P\, e^{-S_F}e^{-S_B},
\label{unq_mf:Zpq}
\end{equation}
where $d\mu_F$ ($d\mu_B$) is the integration measure of the commuting variables related to the fermionic (bosonic) quarks, $S_F$ ($S_B$) is the fermionic (bosonic) action, and $P$ is a prefactor depending on all variables. Evaluating the partially quenched partition function~\eqref{unq_mf:Zpq} at the saddle points, it may happen that the prefactor $P$ is zero at some saddle points, such that these saddles must be disregarded. In this case the replica method can lead to wrong results. In appendix~\ref{sec:wilson}, the occurrence of this phenomenon will be shown in the case of partially quenched theory of massless Wilson fermions.

\vspace{2mm}

To summarize, we have computed the boundaries of the oscillating regions of the eigenvalue density of the baryon number Dirac operator within mean field chiral perturbation theory. As suggested in the example of section~\ref{sec:example} we expect that the unphysical early onset of the average baryon number (see figure~\ref{fig:densities_rho_n}) will disappear due to oscillations inside the regions marked by K on figure~\ref{fig:phases_unq}. In the following section we show, using a RMM, that this is indeed the case.

\section{The oscillations on the microscopic scale}
\label{sec:rmt}

The strong oscillations have a period of order $1/V$ and we therefore need to compute the eigenvalue density on the microscopic scale in order to resolve them. Here we carry out this computation using a random matrix model for the spectrum of the baryon number Dirac operator similar to that of~\cite{Halasz:1997,Halasz:1999}. On the microscopic scale the random matrix model is fully equivalent to chiral perturbation theory, see e.g.~\cite{Akemann:2007} for a review of random matrix theory at nonzero chemical potential.

The RMM partition function with $N_f$ quark flavors, all of mass $m$, in a sector of zero topological charge is given by
\begin{equation}
Z^{N_f,n}_N(\{\mu_f\};m)
=\int\limits_{\mathclap{\C^{N\times N}}}
d\mu(W)\  w(W)\prod_{f=1}^{N_f}\det(D_0(m)+\mu_f),
\label{rmt:Z}
\end{equation}
where $\mu_f$ are the chemical potentials of the different flavors (which at the end all will be set equal to $\mu$), $W$ is a complex $N\times N$ matrix and $w(W)$ is the Gaussian weight function
\begin{equation}
w(W)=\exp[-N\tr WW^\dagger].
\end{equation}
The baryon number Dirac operator, $D_0(m)$, is given by
\begin{equation}
D_0(m)=
\begin{bmatrix}
iW      & m\one_N    \\
m\one_N & iW^\dagger
\end{bmatrix},
\end{equation}
with $\one_N$ the identity matrix. The equivalence between the RMM and $\chi$PT holds in the microscopic limit of QCD with nonzero chemical potential, where the microscopic quantities
\begin{equation}
\hat m\equiv mN\quad\text{and}\quad \hat\mu\equiv\mu\sqrt N
\end{equation}
are kept fixed in the thermodynamic limit, $N\to\infty$, see for example the review \cite{Akemann:2007}. After the standard Hubbard-Stratonovitvich transformation and saddle point integration one sees that the RMM is equivalent to the chiral partition function in microscopic limit, except for an overall factor of $e^{N_fN\mu^2}$ (see e.g.~\cite{Halasz:1999}), under the identifications 
\begin{equation}
\hat m=mN\leftrightarrow m\Sigma V\quad\text{and}\quad \hat\mu^2=\mu^2N\leftrightarrow\mu^2F_\pi^2V.
\end{equation}

It turns out to be a challenging task to write the partition function~\eqref{rmt:Z} as an integral over the joint probability density function of the eigenvalues of $D_0(m)$. For this reason we will focus on the case of massless quarks, where the oscillations are expected to be maximally important and where we can immediately write down the joint probability density function of the eigenvalues of $D_0$.

\subsection{The RMM for Massless Quarks}
\label{sec:rmt_m=0} 

The partition function~\eqref{rmt:Z} simplifies in the massless limit,
\begin{align}
Z^{N_f}_N(\{\mu_f\};m=0)=
\int d\mu(W)\  w(W)\prod_{f=1}^{N_f}&\det(iW+\mu_f)\det(iW^\dagger+\mu_f).
\label{rmt_m=0:Z}
\end{align}
For $N_f=0$ we immediately recognize this as the partition function for the ensemble originally solved by Ginibre~\cite{Ginibre:1965}. If we denote the eigenvalues of $W$ by $\{\eta_k\}$. The eigenvalues of the baryon number Dirac operator $\lambda$ are related to the eigenvalues $\eta$ of $W$ as $\lambda=i\eta$. The partition function can be written as
\begin{equation}
\begin{split}
Z^{N_f=0}_N&=\cN_0\prod_{k=1}^N\int d^2\eta_k\ e^{-N\abs{\eta_k}^2}\abs{\Delta_N(\eta)}^2 \\
&=\cN\prod_{k=1}^N\int d^2\eta_k\ e^{-N\abs{\eta_k}^2} \olddet_{1\leq i,j\leq N} [K_N(\eta_i,\eta_j^\ast)],
\end{split}
\end{equation}
where $\cN_0$ and $\cN$ are normalization constants, $\Delta_N(\eta)$ is the Vandermonde determinant and the kernel $K_N(x,y)$ is defined from the orthogonal polynomials $p_k(x)$ as
\begin{equation}
K_N(x,y)=\sum_{k=0}^{N-1}p_k(x)p_k(y),\quad\text{with}\quad p_k(x)=\frac{\big(\sqrt Nx\big)^k}{\sqrt{k!}}.
\label{rmt_m=0:K}
\end{equation}
The polynomials $p_k$ are orthonormal with respect to the Ginibre weight, i.e.
\begin{equation}
\int_\C d^2\eta e^{-N\abs\eta^2} p_i(\eta)p_j(\eta^\ast)=\frac{\pi\delta_{ij}}{N}.
\end{equation}

Methods to solve matrix models such as~\eqref{rmt_m=0:Z}, where characteristic polynomials are evaluated in a known ensemble, has been given in~\cite{Akemann:2002}. The trick is to use the identity
\begin{equation}
\Delta_{K+L}(x)=\Delta_K(x)\Delta_L(y)\prod_{k=1}^K\prod_{\ell=1}^L(x_k-y_\ell),\quad\text{with}\quad y_\ell=x_{K+\ell}.
\end{equation}
With this trick at hand it is straight forward to carry out the integration in~\eqref{rmt_m=0:Z}. Up to irrelevant factors of normalization, we have
\begin{gather}
\begin{split}
Z^{N_f}_N(\{\mu_f\})&=\cN_0\prod_{k=1}^N\int d^2\lambda_k\ e^{-N\abs{\eta_k}^2}\abs{\Delta_N(\eta)}^2
\prod_{f=1}^{N_f}(i\eta_k+\mu_f)(i\eta_k^\ast+\mu_f) \\
&=\frac{\cN_0}{\Delta_{N_f}(\{i\mu_f\})^2} \prod_{k=1}^N \int dx_kdy_k\Delta_{N+N_f}(x)\Delta_{N+N_f}(y)e^{-Nx_ky_k} \\
&\sim\frac{1}{\Delta_{N_f}(\{i\mu_f\})^2} \prod_{k=1}^N\int dx_kdy_k e^{-Nx_ky_k}
\olddet_{1\leq i,j\leq N+N_f}[K_{N+N_f}(x_i,y_j)] \\
&\sim\frac{1}{\Delta_{N_f}(\{i\mu_f\})^2}\olddet_{1\leq i,j\leq N_f}[K_{N+N_f}(i\mu_i,i\mu_j)],
\label{rmt_m=0:det_K}
\end{split}
\end{gather}
where the kernel $K$ is defined in equation~\eqref{rmt_m=0:K} and $(x,y)$ is defined as
\begin{equation}
x_k=\begin{cases}\eta_k&\text{for}\ k\leq N\\i\mu_{k-N}&\text{for}\ k>N\end{cases}
\quad\text{and}\quad
y_k=\begin{cases}\eta_k^\ast&\text{for}\ k\leq N\\i\mu_{k-N}&\text{for}\ k>N\end{cases}.
\end{equation}
From~\eqref{rmt_m=0:det_K} one sees that the eigenvalue density is given by
\begin{align}
\rho^{N_f}(\eta,\eta^\ast,\{\mu_f\})&=\inner[\Big]{\sum_{k=1}^N \delta^2(\eta_k-\eta)} \nonumber\\
&= \cN'\frac{e^{-N\abs\eta^2}}{Z^{N_f}_N(\{\mu_f\})\Delta_{N_f}(\{i\mu_f\})^2}
\begin{vmatrix}
K_{N+N_f}(\eta,\eta^\ast)       & \cdots & K_{N+N_f}(\eta,i\mu_{N_f})   \\
\vdots                                &        & \vdots                        \\
K_{N+N_f}(i\mu_{N_f},\eta^\ast)    & \cdots & K_{N+N_f}(i\mu_{N_f},i\mu_{N_f})
\end{vmatrix},
\label{rmt_m=0:rho}
\end{align}
where $\cN'$ is a normalization constant to be canceled by the normalization constant in $Z^{N_f}_N(\{\mu_f\})$. The spectral density~\eqref{rmt_m=0:rho} is written in terms of $\eta$, the eigenvalues of $W$, but as mentioned previously these eigenvalues are related to the eigenvalues of the baryon number Dirac operator by $\lambda=i\eta$. The spectral densities for $\lambda$ and $\eta$ is related by a rotation by $\pi$ in the complex plane, such that baryon spectral density is easily obtained from~\eqref{rmt_m=0:rho}. Some calculations are more compact in terms of the $\eta$ eigenvalues, and we will for this reason keep on writing expressions in terms of $\eta$ in the following section.

The complex structure of the spectral density~\eqref{rmt_m=0:rho} introduces oscillations with a microscopic period. In the following section we show that the oscillations are bounded within the region predicted in section~\ref{sec:mf} by mean field $\chi$PT. Futhermore we will explicitly demonstrate the crucial role of the oscillations.

\subsection{The average baryon density from the oscillating eigenvalue density}
\label{sec:N_f=1}

\begin{figure}[htbp]
\begin{minipage}[t]{.999\textwidth}
\includegraphics[width=10cm]{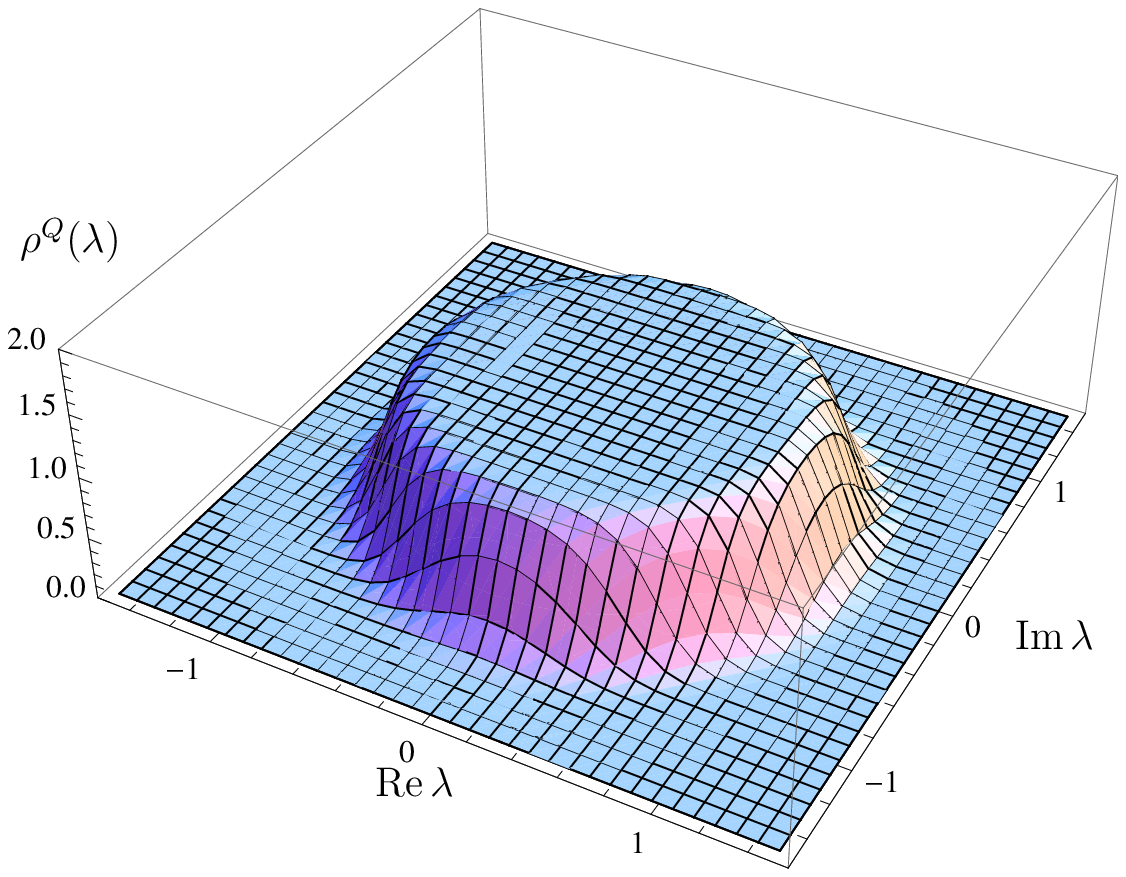} 
\includegraphics[width=10cm]{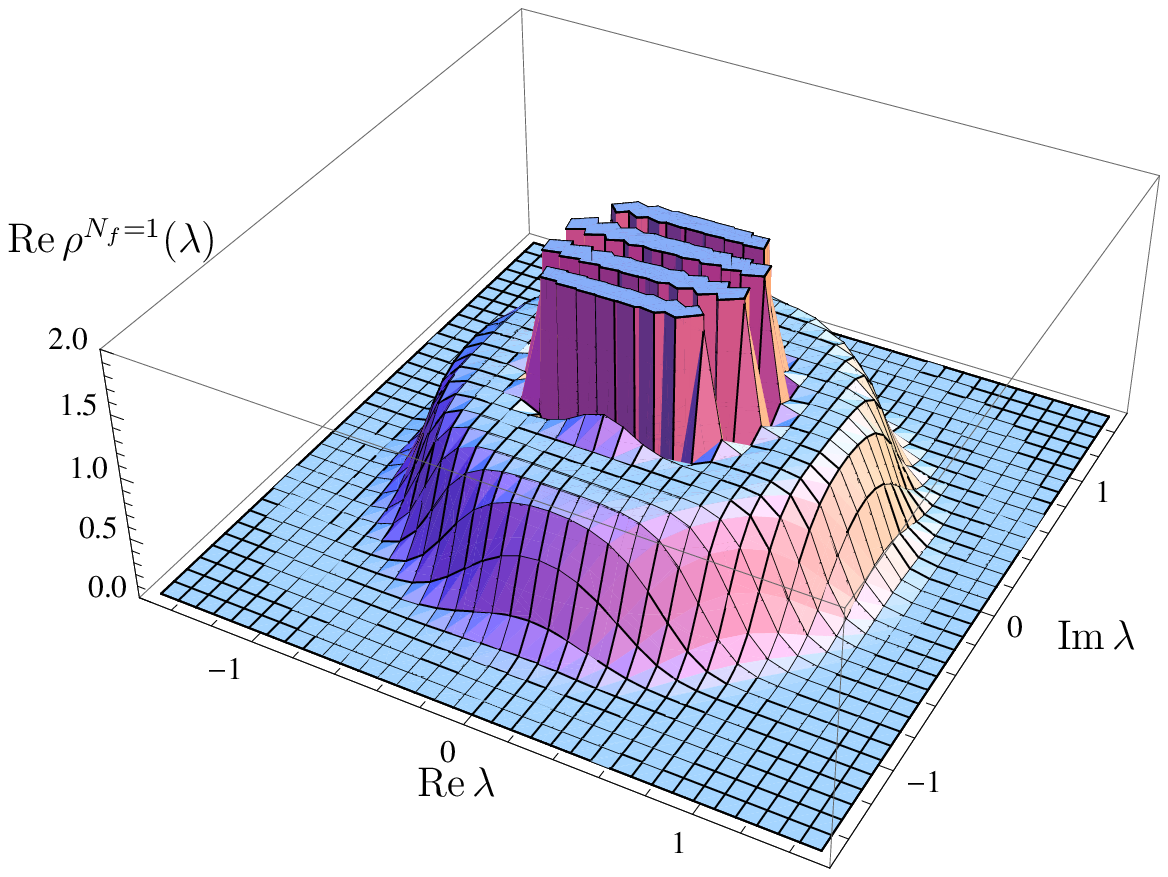}
\end{minipage}
\caption{\textbf{Top panel:} The quenched spectral density with $N=36$. This density is identical to the Ginibre distribution. \textbf{Bottom panel:} The $N_f=1$ spectral density with $N=36$ and $\mu=0.45$. One sees how unquenching introduces a strongly oscillating region. The amplitude of the oscillations grows exponentially with $N$ and the period is proportional to $1/N$. The peaks of the oscillations has been cut; the maximum amplitude is more than a hundred orders of magnitude larger than the scale displayed.}
\label{fig:spectral_dens.}
\end{figure}

For a chemical potential less than a third of the nucleon mass, the theory is dominated by pions and the partition function is therefore independent of the chemical potential at low temperature. In this section we will show how the $\mu$-independence enters the RMM. Exactly as in the example of section \ref{sec:example}, the baryon number Dirac spectrum has a strong dependence on $\mu$ (cf. figure~\ref{fig:spectral_dens.}), even though the average baryon density does not. As for the calculation for the chiral condensate~\cite{Akemann:2004,Osborn:2008b}, the $\mu$-independence of the observable is obtained from a strongly oscillating region of the eigenvalue density.

As already mentioned, the quenched spectral density (seen in the top panel of figure~\ref{fig:spectral_dens.}) is the Ginibre distribution,
\begin{equation}
\rho^Q(\eta,\eta^\ast)=e^{-N\abs\eta^2}K_N(\eta,\eta^\ast)=\frac{\Gamma(N,N\abs\eta^2)}{\Gamma(N)}.
\end{equation}
The one flavor partition function~\eqref{rmt_m=0:det_K}, is given by the kernel,
\begin{equation}
Z^{N_f=1}_N(\mu)=K_{N+1}(i\mu,i\mu),
\end{equation}
and agrees with the result derived in~\cite{Halasz:1997} by a different method. The one flavor spectral density is given by~\eqref{rmt_m=0:rho},
\begin{align}
\rho^{N_f=1}(\eta,\eta^\ast,\mu)=2e^{-N\abs\eta^2}
\Big[K_{N+1}(\eta,\eta^\ast)-\frac{K_{N+1}(\eta,i\mu)K_{N+1}(i\mu,\eta^\ast)}{K_{N+1}(i\mu,i\mu)}\Big].
\label{rmt_m=0:rho_Nf=1}
\end{align}
One can easily check that the spectral density~\eqref{rmt_m=0:rho_Nf=1} contain a circular region of radius $\mu$ with strong oscillation, exactly as predicted in section~\ref{sec:mf}. The lower panel of figure~\ref{fig:spectral_dens.} illustrates this.

Here we will show that the oscillations due to the second term in~\eqref{rmt_m=0:rho_Nf=1} exactly cancel the unphysical early onset of the quenched baryon density.
The baryon density is defined from the partition function as
\begin{equation}
Vn_B^{N_f=1}(\mu)=\frac{d}{d\mu}\log [Z^{N_f=1}_N(\mu)]=\frac{dK_{N+1}(i\mu,i\mu)/d\mu}{K_{N+1}(i\mu,i\mu)}.
\label{rmt_m=0:n_B}
\end{equation}
We will now compute the average baryon number starting from the eigenvalue density, \eqref{rmt_m=0:rho_Nf=1} using
\begin{equation}
Vn_B^{N_f=1}(\mu)=\int_\C d^2\eta\frac{1}{i\eta+\mu}\rho^{N_f=1}(\eta,\eta^\ast,\mu),
\end{equation}
and show that the oscillating nature is essential in order to get agreement with~\eqref{rmt_m=0:n_B}.

After addition and subtraction of a kernel $K_{N+1}(i\mu,\eta^\ast)$ the one flavor spectral density given in~\eqref{rmt_m=0:rho_Nf=1} can be written as
\begin{multline}
\tfrac{1}{2}\rho^{N_f=1}(\eta,\eta^\ast,\mu)=
w(\eta,\eta^\ast)\big[K_{N+1}(\eta,\eta^\ast)-K_{N+1}(i\mu,\eta^\ast)\big] \\
-w(\eta,\eta^\ast)
\Big[\frac{K_{N+1}(i\mu,\eta^\ast)}{K_{N+1}(i\mu,i\mu)}(K_{N+1}(\eta,i\mu)-K_{N+1}(i\mu,i\mu))\Big],
\end{multline}
with the weight $w(\eta,\eta^\ast)=e^{-N\eta\eta^\ast}$. Integration over the first term of the spectral density,
\begin{equation}
\int d^2\eta \frac{w(\eta,\eta^\ast)}{i\eta+\mu}\big[K_{N+1}(\eta,\eta^\ast)-K_{N+1}(i\mu,\eta^\ast)\big]
=-i\int d^2\eta w(\eta,\eta^\ast)\sum_{k=0}^N\frac{p_k(\eta)-p_k(i\mu)}{\eta-i\mu}p_k(\eta^\ast),
\end{equation}
is zero due to orthogonality. In evaluation of the nonzero integral we will exploit that the eigenvalues of the baryon number Dirac operator come in pairs $(i\eta,i\eta^\ast)$ combined with the fact that the weight $w(x,y)$ and the kernel $K_n(x,y)$ only depend on the product $xy$ to write,
\begin{equation}
\begin{split}
Vn_B^{N_f=1}(\mu)=
&+i\int d^2\eta w(\eta,\eta^\ast)
   \frac{K_{N+1}(i\mu,\eta^\ast)}{K_{N+1}(i\mu,i\mu)}\frac{K_{N+1}(\eta,i\mu)-K_{N+1}(i\mu,i\mu)}{\eta-i\mu} \\
&-i\int d^2\eta w(\eta,\eta^\ast)
   \frac{K_{N+1}(\eta,-i\mu)}{K_{N+1}(i\mu,i\mu)}\frac{K_{N+1}(-i\mu,\eta^\ast)-K_{N+1}(i\mu,i\mu)}{\eta^\ast+i\mu}.
\end{split}
\end{equation}
The first integral only has a contribution from $\eta=i\mu$ and the second only from $\eta^\ast=-i\mu$. The reason for this is that $K_{N+1}(i\mu,\eta^\ast)$ and $K_{N+1}(\eta,-i\mu)$ are reproducing kernels in the space of polynomials of order less than $N$, and thus in a way are equivalent to delta functions $\delta^2(\eta-i\mu)$ and $\delta^2(\eta^\ast+i\mu)$, respectively. Using this we see that the average baryon number can be written as
\begin{equation}
\begin{split}
Vn_B^{N_f=1}(\mu)=
&+i\frac{\lim_{\eta\to i\mu}}{K_{N+1}(i\mu,i\mu)}\frac{K_{N+1}(\eta,i\mu)-K_{N+1}(i\mu,i\mu)}{\eta-i\mu} \\
&-i\frac{\lim_{\eta^\ast\to-i\mu}}{K_{N+1}(i\mu,i\mu)}\frac{K_{N+1}
(-i\mu,\eta^\ast)-K_{N+1}(i\mu,i\mu)}{\eta^\ast+i\mu} \\
=&+\frac{dK_{N+1}(i\mu,i\mu)/d\mu}{K_{N+1}(i\mu,i\mu)}.
\end{split}
\end{equation}
This is precisely the result we found in~\eqref{rmt_m=0:n_B}, and the computation demonstrates directly the crucial role of the strong oscillations. Note that this result is true for all values of $N$. It therefore automatically applies in the microscopic limit, where the RMM is equivalent to $\chi$PT. 

\vspace{2mm}
\begin{figure}[htbp]
\includegraphics[width=7cm]{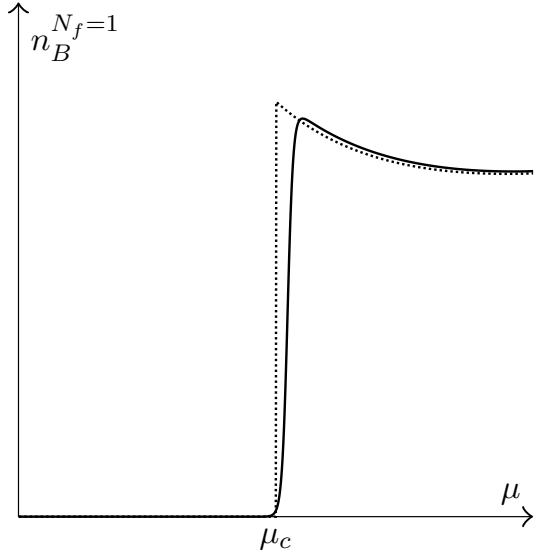}
\caption{Plot of the baryon density in the RMM for $N=36$ and $m=0$ (which implies $m_\pi=0$) as a function of the chemical potential. The dotted line indicates the large $N$ limit. The baryon density is seen to be zero below the phase transition at $\mu_c=0.527\ldots$ \cite{Stephanov:1996} which is the RMM analog of a third of the nucleon mass.}
\label{fig:n_B}
\end{figure}
As mentioned above the RMM partition function and the $\chi$PT partition function differ by a trivial overall factor $e^{N_f\hat\mu^2}$ in the microscopic limit. If one want show that the $\mu$-independence of average baryon density one need to multiply this trivial factor on the RMM partition function before further calculation. The one flavor baryon density is therefore given by
\begin{equation}
Vn_B^{N_f=1}(\mu)=\frac{d}{d\mu}\log [Z^{N_f=1}_N(\mu)e^{N\mu^2}]=\frac{dK_{N+1}(i\mu,i\mu)/d\mu}{K_{N+1}(i\mu,i\mu)}+2N\mu.
\label{rmt_m=0:n_B+}
\end{equation}
To evaluate the average baryon density~\eqref{rmt_m=0:n_B+} in the large $N$ microscopic limit, we write the partition function as
\begin{equation}
Z^{N_f=1}_N(\mu)e^{N\mu^2}=\frac{\Gamma(N+1,-N\mu^2)}{\Gamma(N+1)}=\frac{1}{(N-1)!}\int_{-\mu^2}^\infty dt\, e^{N(\log t-t)}.	
\end{equation}
This integral can be evaluated with saddle point integration. There are two local maxima: One at $t=1$ and one at $t=-\mu^2$. The critical value of the chemical potential $\mu_c$, which decides which maximum is the global one, is given by~\cite{Stephanov:1996}
\begin{equation}
\mu_c^2+\log\mu_c^2+1=0.
\end{equation}
For $\abs\mu<\mu_c$ the integral is dominated by $t=1$ and partition function is independent of $\mu$, hence the average baryon number is zero. For $\abs\mu>\mu_c$ the average baryon number is given by $Vn_B^{N_f=1}\approx 2N(\mu+\mu^{-1})$. This is shown on figure~\ref{fig:n_B}. The critical $\mu_c$ is thus the RMM analog of a third of the nucleon mass \cite{Halasz:1998qr}. Beyond the critical chemical potential the system is no longer dominated by the Goldstone bosons and the equivalence of the RMM and $\chi$PT breaks down.

\vspace{3mm}

Finally, let us note that the spectrum of the baryon number Dirac operator 
has a $2\pi nT$-periodicity along the imaginary axis, where $T$ is the 
temperature. This is a manifestation
of the fact that if $\psi$ is an eigenfunction of $D_0(m)$ with
eigenvalue $\lambda$, then $e^{i\omega_n x_0}\psi$ is an
eigenfunction with eigenvalue $\lambda+i\omega_n$~\cite{Cohen:2003}.
From the boundary condition of $\psi$ in the 
time direction it follows that $\omega_n=2\pi inT$, where $n$ is an
integer \cite{Gibbs:unp}. This periodicity is related to the periodicity 
of imaginary 
chemical potential, see~\cite{Roberge:1986mm}, but is not seen in the
microscopic limit, since the eigenvalues scale as $\lambda\sim T^2$.

\section{conclusions}
\label{sec:con}

In QCD with nonzero chemical potential the sign problem manifest itself in the physical observables. This is true for the average baryon number as well as for the chiral condensate. If one neglects the phase factor of the fermion determinant the baryon density will have an unphysical early onset at $\mu=m_\pi/2$ and the chiral condensate will rotate into a pion condenssate.

In this paper the correct physical behavior of the unquenched average baryon number has been linked to strong oscillations of the eigenvalue density of the baryon number Dirac operator. The oscillations have a period on the microscopic scale and were shown to be dominant within bounded regions of the complex eigenvalue plane. The mechanism which links the oscillations of the baryon number Dirac spectrum to the baryon number density is in exact correspondence with the mechanism between the chiral condensate and the Dirac spectrum. This shows the general nature of this mechanism, when a sign problem is present. 
The boundaries of the oscillating regions of the baryon number Dirac spectrum where computed within mean field chiral perturbation theory and using a random matrix model it was shown exactly how the oscillations are responsible for the $\mu$-independence of the average baryon density.

In conclusion, we have shown that the mechanism which links physical observables in unquenched QCD to strong oscillations of the corresponding eigenvalue density is not restricted to the chiral condensate; it also holds for the average baryon number. This solves the Sliver Blaze problem \cite{Cohen:2003} for the baryon number Dirac operator. The oscillations have a period of order $1/V$ and an amplitude that grows exponentially with $V$. Moreover, in both cases one must integrate over at least $V$ periods of the oscillations to approach the correct physical behavior. This demonstrates how severe the sign problem is for $\mu>m_\pi/2$, see also \cite{Lombardo:2009rt}. 

It is an appealing mathematical challenge to generalize the computation carried out here within the random matrix framework to the case of a nonzero quark mass. The direct supersymmetric computation of the microscopic spectral density is extremely demanding in this case, but perhaps it can be simplified through the use of integrable structures as in \cite{Splittorff:2002,Splittorff:2004}.

It would be interesting to extend this study to the 1dQCD baryon number Dirac operator. Finally, it would also be most interesting to study the results of the present paper within the approach of \cite{Hanada:2012es}. 

\paragraph*{Acknowledgments:}
We wish to thank G. Akemann, P. H. Damgaard, T. Guhr, M. Kieburg and J.J.M. Verbaarschot as well as the participants of the ZiF workshop ’Random Matrix Theory and Applications in Theoretical Sciences’ for useful discussions. The work of K.S. was supported by the Sapere Aude program of The Danish Council for Independent Research.

\appendix

\section{Partially Quenched Wilson Fermions}
\label{sec:wilson}

In section~\ref{sec:mf} it was mentioned that saddle point integration within the replica method may give saddle points that should be neglected. This happens when the corresponding saddle point in the partially quenced theory has a prefactor identical to zero. In this appendix we show an example of this phenomenon in the effective theory of QCD with Wilson fermions.

The partition function describing QCD with $N_f$ flavors of Wilson fermions in a sector with zero topological charge is
\begin{equation}
Z_{N_f}(M,a)=\int_{U(N_f)}dU e^{-S(U,M,a)}.
\end{equation}
In the microscopic limit the action is given as
\begin{multline}
S(U,M,a)=-\frac{1}{2}\Sigma V\tr M(U+U^\dagger) \\
-a^2W_6V[\tr(U+U^\dagger)]^2-a^2W_7V[\tr(U-U^\dagger)]^2-a^2W_8V\tr(U^2+U^{\dagger2}),
\end{multline}
where $M$ is the mass matrix, $a$ is the lattice spacing, $\Sigma$ is the chiral condensate and $W_i$ are low energy constants which determine the leading order discretization errors of Wilson fermions~\cite{Sharpe+}. Here we will focus on the special case where $W_6=W_7=0$. The scaled variables, $\hat M=M\Sigma V$ and $\hat a^2_8=a^2W_8V$, are kept at order unity in the microscopic limit. One finds the partially quenched partition function~\cite{Damgaard:2010,Akemann:2010,Splittorff:2011bj}
\begin{equation}
Z_{N_f+1\vert 1}(M,a_8)=\int dU \,e^{\frac{i}{2}\str\hat M(U-U^{-1})+\hat a^2_8\str(U^2+U^{-2})}.
\end{equation}
The integration is over the maximum Riemannian submanifold of $Gl(N_f+1\vert1)$. For $N_f=0$ we can parameterize the graded manifold as~\cite{Damgaard:1999}
\begin{equation}
U=\begin{pmatrix}e^{i\theta}&0\\0&e^s\end{pmatrix}\exp\begin{pmatrix}0&\alpha\\\beta&0\end{pmatrix},
\end{equation}
where $\alpha$ and $\beta$ are Grassmann variables. Setting $M=m\one$ and integrating out the Grassmann variables gives
\begin{equation}
Z_{1\vert 1}(m,a_8)=\frac{1}{2\pi}\int_{-\infty}^\infty ds \int_{-\pi}^\pi d\theta\, 
P(m,a,\theta,s)e^{-S_f(m,a_8,\theta)}e^{-S_b(m,a_8,s)},
\label{wilson:Z}
\end{equation}
with
\begin{align}
S_f(m,a_8,\theta)=&{}+\hat m\sin\theta+2\hat a_8^2\cos2\theta, \\
S_b(m,a_8,\theta)=&{}+i\hat m\sinh s  -2\hat a_8^2\cosh2s, \\
P(m,a_8,\theta,s)=&{}-\frac{\hat m}{2}\sin\theta+i\frac{\hat m}{2}\sinh s \nn \\
&+2\hat a^2_8(\cos2\theta+\cosh2s+2\cos\theta\cosh s-2i\sin\theta\sinh s).
\end{align}
Numerically one can easily verify that the partition function~\eqref{wilson:Z} equals one as it should, independently of $\hat{m}$ and $\hat{a}_8$, since we work at equal fermion and boson sources.  Here we want to calculate~\eqref{wilson:Z} with saddle point integration. The saddle points are given by
\begin{subequations}
\begin{align}
s&=0,\ \cos\theta=0 \label{wilson:saddle1} \\
s&=0,\ \sin\theta=-\frac{\hat{m}^2}{8\hat{a}_8^2}. \label{wilson:saddle2}
\end{align}
\end{subequations}
In general both saddle points will contribute to the partition function, but note that the prefactor $P$ evaluated at the saddle point~\eqref{wilson:saddle1} is
\begin{equation}
P(\hat{m},\hat{a}_8,\theta=-\tfrac{\pi}{2},s=0)=\frac{\hat{m}}{2},
\end{equation}
such that the prefactor becomes zero in the massless limit. Calculating the partition function~\eqref{wilson:Z} in massless limit with saddle point integration one only needs to include the saddle point~\eqref{wilson:saddle2},
\begin{equation}
Z^\nu_{1\vert 1}(m=0,a_8)=
\frac{P(a,\theta,s)e^{-S_f(a_8,\theta)}e^{-S_b(a_8,s)}}{\abs{S_f''(a_8,\theta)S_b''(a_8,s)}}
\Bigg\vert_{\theta=s=0}=1.
\end{equation}
This behavior is in complete analogue to the behavior of the massless theory mentioned in section~\ref{sec:mf}. In such cases the mean field replica method can lead to wrong results.

\end{document}